\documentstyle[aps,epsf,prl,multicol]{revtex}

\begin{document}
\draft
\preprint{HKBU-CNS-9904}
\title
{Heat conduction in one dimensional nonintegrable systems}
\author{Bambi Hu$^{1,2}$, Baowen Li$^{1,3}$
\footnote{To whom correspondence should be addressed. Email: 
bwli@hkbu.edu.hk}, 
and Hong Zhao$^{1,4}$
}
\address{
$^1$ Department of Physics and Centre for Nonlinear Studies, Hong 
Kong Baptist University, China\\
$^2$ Department of Physics, University of Houston, Houston TX 77204-5506\\ 
$^3$ Department of Physics, National University of Singapore, 
119260 Singapore\\
$^4$ Department of Physics, Lanzhou University, 730000 Lanzhou, China
}
\date{Received 17 June 1999} 
\maketitle

\begin{abstract} 
Two classes of 1D nonintegrable systems 
represented by the Fermi-Pasta-Ulam (FPU) model and the discrete 
$\phi^4$ model are studied to seek a generic
mechanism of energy transport in microscopic level
sustaining  macroscopic behaviors. The results enable us to understand 
why the class 
represented by  the $\phi^4$ model has a normal thermal conductivity 
and the class represented by 
the FPU model does not even though the temperature gradient can be 
established. 

\end{abstract} 


\begin{multicols}{2}

Heat conduction in one-dimensional (1D) nonintegrable Hamiltonian systems 
is a vivid example for studying microscopic origin of the
macroscopic irreversibility in terms of deterministic chaos. It is
one of the oldest but a rather fundamental problem in nonequilibrium
statistical mechanics\cite{chaos98}. Intended to understand the underlying
mechanism of the Fourier heat conduction law, the study of heat conduction
has attracted increasing attention in recent years
\cite{Lebowitz78,Casati84,FPU,HLZ98,FHLZ98,Hatano99,HLZ99,Dhar99,Casati99,TLH99}. 

Based on previous studies, we can classify the 1D lattices into three
categories. The first one consists of integrable systems such as the harmonic
chain. It was rigorously shown \cite{Rieder} that, in this category no
temperature gradient can be formed, and the thermal conductivity is
divergent. The second category includes a number of nonintegrable systems
such as the Lorentz gas model\cite{Lebowitz78,Casati99}, the ding-a-ling
and alike models\cite{Casati84}, and the Frenkel-Kontorova (FK)
model\cite{HLZ98} and etc.. In this category, the heat current is
proportional to $N^{-1}$ and the temperature gradient $dT/dx\sim
N^{-1}$, thus the thermal conductivity $\kappa$ is a constant independent
of system size $N$. The Fourier heat conduction law ($J = -\kappa dT/dx
$) is justified.  The third category also includes some nonintegrable
systems such as the FPU\cite{FPU,FHLZ98} chain, the diatomic Toda
chain\cite{Hatano99}, the (mass) disorder chain\cite{HLZ99}, and the 
Heisenberg spin chain\cite{Dhar99} and so on.
In this category, although the temperature gradient can be set up with
$dT/dx \sim N^{-1}$, the heat current is proportional to
$N^{\alpha-1}$ with $\alpha\sim 0.43$, the thermal conductivity $\kappa
\sim N^{\alpha}$ which is divergent as one goes to the thermodynamic limit
$N\to\infty$. 

These facts suggest that the {\it nonintegrability is necessary to have a
temperature gradient, but it is not sufficient to guarantee the normal
thermal conductivity} in a 1D lattice. This picture brings us to ask two
questions of fundamental importance: (i) Why do some nonintegrable systems
have normal thermal conductivity, while the others fail? (ii) How can the
temperature gradient be established in those nonintegrable systems
having divergent thermal conductivity? 

The reason for the 
divergent thermal conductivity in integrable system  is that the energy 
transports freely
along the chain without any loss
so that no temperature gradient can 
be established. The set up of temperature gradient in 
nonintegrable systems implies the existence of 
scattering. However, 
the different heat conduction behavior in the two
categories of nonintegrable systems indicates that the underlying 
mechanism must be different. To get the point, let's 
write the Hamiltonian of a generic 1D lattice as
\begin{equation}
H = \sum_i H_i, \qquad H_i=\frac{p_{i}^{2}}{2} + V(x_{i-1},x_i) + 
U(x_i),
\label{Ham}
\end{equation}
where $V(x_{i-1},x_i)$ stands for the interaction potential of the 
nearest-neighbor particle and $U(x_i)$ is an external (or on-site) 
potential. The origin of external potential in real 
physical systems varies from model to model. For instance, in the 
FK model\cite{HLZ98} the external potential is the 
interaction of the adsorbed atoms with the crystal surface.
It is $U(x)$ that 
distinguishs from the two categories of nonintegrable 
lattices. $U(x)$ vanishes
in all 1D lattices having divergent thermal 
conductivity. We are thus convinced to conclude 
that the {\it external potential plays a determinant role for 
normal thermal conduction}. 

In this paper we would like to study the scattering mechanism and
the role of the external potential in heat conduction in the two 
categories of nonintegrable
systems. For this purpose, we choose two representatives from these two
categories, i.e. the discrete $\phi^4$ model (see, e.g.  Ref.
\cite{Aubry96}) and the FPU model. Both models are the simplest anharmonic
approximation of a monoatomic solid. In the $\phi^4$ model, $V$ takes the
harmonic form, and the external potential $ U(x) = m x^2/2 + \beta x^4/4$,
with $m$ fixed to be zero in this paper.  In the FPU model, $U$ vanishes
and $V$ takes the anharmonic form of $(x_i-x_{i-1})^2/2
+\beta(x_i-x_{i-1})^4/4$, and $\beta=1$ throughout this paper. In the case of
$\beta=0$, the FPU model reduces to the harmonic chain. 

In our numerical simulations the Nos\'e-Hoover thermostates \cite{NH85}
are put on the first and the last particles, keeping them at 
temperature $T_+$ and $T_-$, respectively. The motion of these two 
particles are governed by
\begin{eqnarray}
\ddot{x}_1 = -\zeta_+ \dot{x}_1 + f_1 - f_2, \quad
\dot{\zeta}_+= \dot{x}_1^2/T_+ -1\nonumber\\
\ddot{x}_N = -\zeta_- \dot{x}_N + f_N - f_{N+1}, \quad
\dot{\zeta}_-= \dot{x}_1^2/T_- -1.
\label{eqm2}
\end{eqnarray}
where $f_i = -(V' + U')$ is the force acting on the $i$'th particle. The 
equation of motion of other particle is $\ddot{x}_i = 
f_i - f_{i+1}$. The eighth-order Runge-Kutta algorithm was 
used. All computations are carried out in double precision. 
Usually the stationary state set in after  $10^7$ time units. We 
should point out that we have performed computations by using other 
types of thermostate, and no qualitative difference has 
been found.

\begin{figure}
\epsfxsize=8cm
\epsfbox{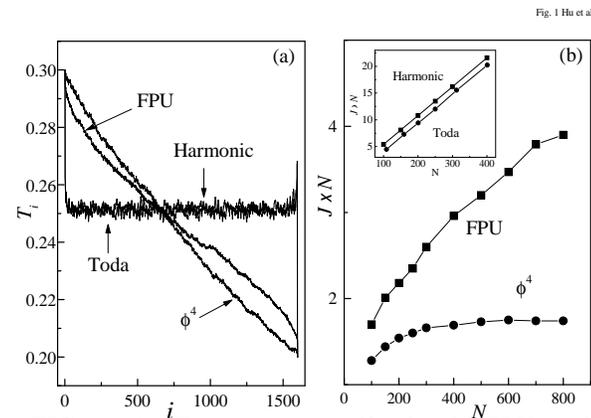}
\vspace{-.5cm}
\narrowtext
\caption{
(a) Temperature profiles for the FPU model, the $\phi^4$ model,  the 
harmonic and the monoatomic Toda model. (b) The quantity  
$J\times N$ versus the system size $N$ for the FPU (solid square) and the 
$\phi^4$ (solid circle). That of  
the harmonic and the monoatomic Toda are shown in the inset of (b). The 
lines in (b) and its inset are drawn for guiding the eye. $T_{+}=0.3$, 
$T_-=0.2$. } 
\end{figure}
 
Fig. 1(a) shows temperature profiles. In all 
nonintegrable systems, the temperature scales as $T=T(i/N)$. 
However, in the FPU case there is a singular
behavior near the two ends, which is a typical character
of 1D nonlinear lattices having divergent thermal conductivity. In the
same figure we also show the temperature profiles for two integrable
lattices: the harmonic and the monoatomic Toda models.  In these two cases
no temperature gradient could be set up and the stationary state
corresponds to
$T=(T_{+}+T_-)/2$, which is consistent with the rigorous result
\cite{Rieder}. 

In Fig. 1(b), we plot the quantity $J\times N$ versus $N$
for the FPU model and the
$\phi^4$ model. The inset shows the same quantity for
the harmonic chain and the monoatomic Toda chain. 
The local heat flux is defined by 
$J_i =\dot{x}_i \frac{\partial V}{\partial x_{i+1}}$. We found that  
when the system reaches a stationary state, the time 
average $\langle J_i\rangle$ is site independent, it is denoted as 
$J$. For the harmonic chain and the monoatomic Toda 
chain $J\times N$ is expected to be proportional to $N$ since 
$J$ is $N$-independent. This is indeed the case as illustrated in the 
inset. In both the FPU and 
the $\phi^4$ models $dT/dx$ is proportional to
$-1/N$, the thermal conductivity $\kappa =-J/(dT/dx) \propto 
J\times N$. Fig. 1(b) tells us
that, in contrast to the integrable systems and the FPU model,  
heat conduction in the $\phi^4$ model obeys the Fourier law. 

\begin{figure}
\epsfxsize=8cm
\epsfbox{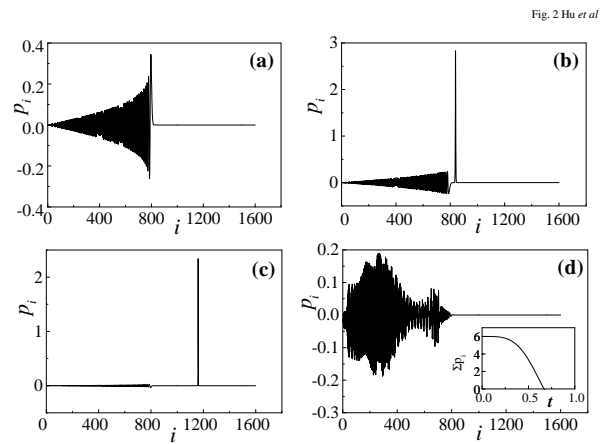}
\narrowtext
\caption{
The momentum excitation in different lattices. (a) the harmonic chain; 
(b) the monoatomic Toda chain; (c) the FPU model; and (d) the $\phi^4$ 
model. Inset in 
(d) illustrates the time evolution of the total momentum of the $\phi^4$ 
chain.} 
\end{figure}

The heat current $J$ in all nonintegrable systems decreases as
the system size $N$ is increased ($J\sim N^{\alpha -1}, 0<\alpha<1$). To 
clarify the underlying mechanism we decompose the
interaction of the thermostat into a series of kicks and study the
transport of a single kick along the chain. A free boundary condition is
used in our calculation, but we should stress that the results do not
depend on the types of boundary condition. In Fig. 2 we plot $p_i$ versus
$i$ after a long time ($t=800)$ for four lattices: the harmonic chain (a);
the monoatomic Toda chain (b); the FPU model (c); and the $\phi^4$ model
(d). The amplitude of the wave profile in the harmonic chain decreases
continuously with time, but the global profile keeps unchanged. In both
the FPU and the Toda models, we observe a solitary wave separates from the 
long tail. Initially this wave front is connected with other low amplitude
excitations. After a certain time, this wave front, moves faster,
separates
from the tail and goes forward and  keeps its amplitude. In the mean
time the tails behind it evolve in the same way as that in
the harmonic chain. In the $\phi^4$ model, the head part of the profile
becomes weaker and weaker. The reason is that in the first three cases (a-c)
both the total energy and the total momentum are conserved, whereas in the
$\phi^4$ model the momentum conservation breaks down due to the
external potential. The inset in Fig. 2(d) shows that the
total momentum in the $\phi^4$ model decreases at least exponentially with
time. The decay of the momentum with time indicates a loss of correlation.
It is thus reasonable to envisage the energy transport along the $\phi^4$
chain as a random walk-like scattering.

The solitary waves in the FPU chain exchange energy and momentum when
colliding with each other. It causes the energy loss, and the heat
current decreases when the system size is increased.  To show this, we
start two excitations at the two ends of the chain with different
momentum, one moves to the right and another left. Let $p_1=6$, $p_N=3$
and $p_i=0, i \neq 1$ and $ N $ as our initial excitations. We calculate
the momentums of the solitary waves (by simply summing up momentums of
several lattices around the peaks) and investigate its change before and
after the interaction. We find that the bigger one generally transfers
part of its momentum and energy to the smaller one, as is shown in Fig.
3(a). The collision takes place at $t=850$, where a peak is shown up.

\begin{figure}
\epsfxsize=8cm
\epsfbox{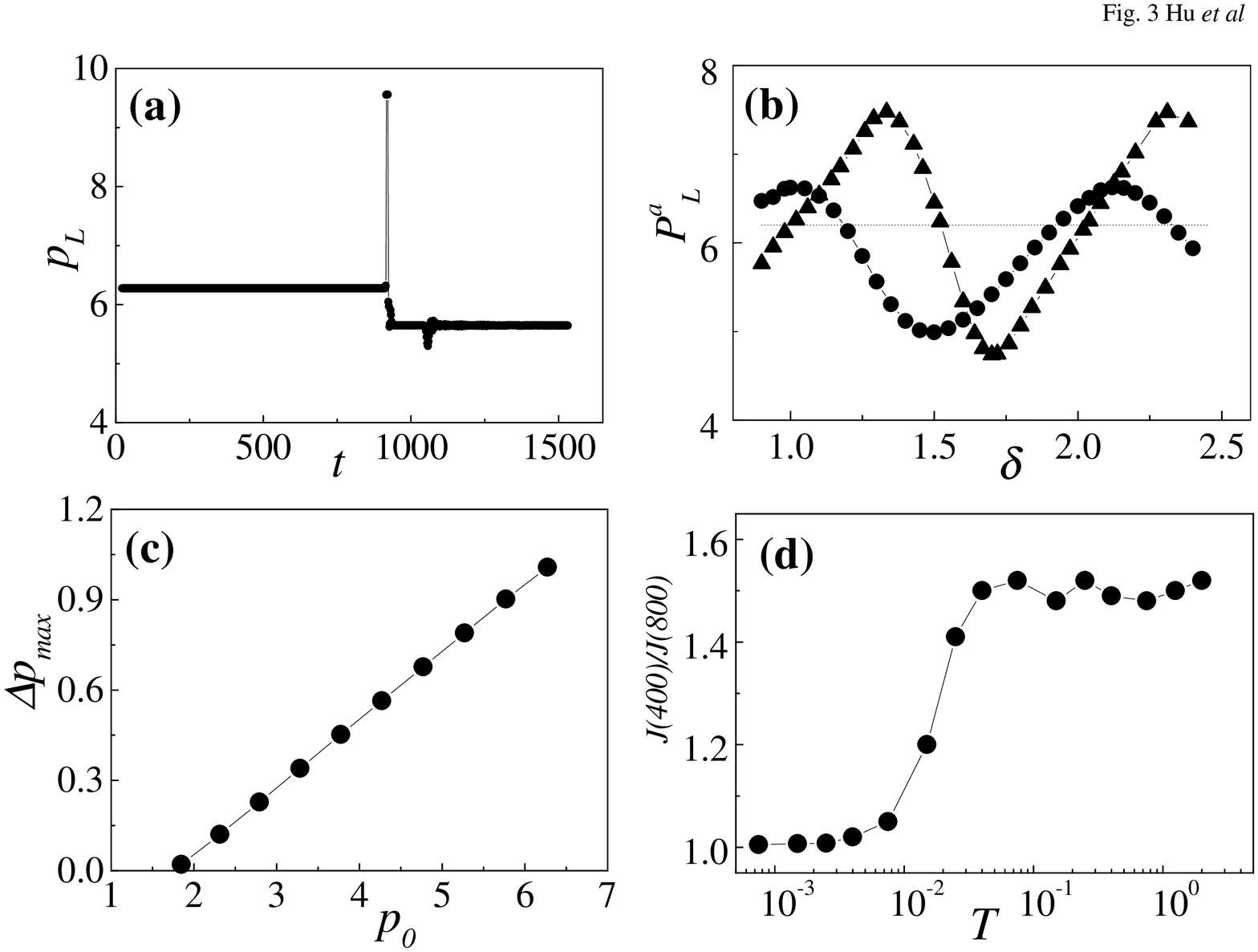}
\narrowtext
\caption{(a) The time dependence of the momentum of a solitary wave. (b) 
The momentum (after collision) versus the phase. Solid triangle represents 
the results of the case in which both left and right solitary waves 
have the same initial momentum. Solid circle represents the case of 
the solitary waves having different initial momentum (see text for more 
in detail). The horizontal line is the momentum before collision. (c) The 
maximal momentum gain $\Delta p_{max}$ versus initial 
momentum $p_0$. (d) The ratio of the heat flux of $J(400)/J(800)$ versus 
the average temperature $T=(T_++T_-)/2$ in the semi-logarithmic scale. 
The results shown here are for the FPU model.} 
\end{figure}

Moreover, the interaction between solitary waves is found to depend
closely on a ``phase'' difference. Here the ``phase'' difference is
defined as a time lag between the excitations of two solitary waves. For
instance, if we excite a solitary wave from the left end at time $t$, and
another one from the right end at time $t+\delta$, then $\delta$ is the
``phase'' difference. These two solitary waves, traveling through the
chain
in opposite directions, will collide with each other after a certain time. 
Although the physical meaning of the ``phase'' is not obvious, it
is an important and good quantity to describe the
interaction. We show $p^a_L$ versus $\delta$ for two different kinds of
collision in Fig 3(b), where $p^a_L$ is the momentum of the solitary wave
from the left after collision. In the first case
both left and right solitary waves have the same initial momentum
$p_L=p_R=6.27$, which is excited by an initial condition of $p_1=p_N=6$
and $p_i=0$ for other $i$.  In the second case, the left one has $p_L=6.27$
and the right one $p_R=3.28$, excited by an initial condition of $p_1=6$,
$p_N=3$ and $p_i=0$ for $i \neq $ $1$ and $N$. The figure shows that 
$P_L^a$ depends on the ``phase'' sinusoidally. 

Other interesting features of the collision of solitary waves are shown in
Fig. 3(c), where we plot the maximum momentum gain $\Delta p_{max}$ versus
the initial momentum $p_0$ for the FPU model. $\Delta p_{max}$ is measured
by subtracting the initial momentum $p_0$ from the maximum $p_L$ in Fig.
3b. First of all, this picture tells us that the exchange of momentum and
energy depends on the initial momentum and energy. Secondly, there exists
a critical momentum below which no energy exchange can take place.  The
critical $p^c_0\sim1.8$ is clearly seen in the figure. For $p_0 < p^c_0$,
$\Delta p_{max}$ is zero. This result is very significant, it indicates
that there exists a threshold for the solitary wave interaction, below
this threshold the interaction ceases, i.e. no momentum and energy
exchanges between the solitary waves. A direct consequence of this fact is
the existence of a threshold temperature below which the FPU chain should
behave like a harmonic chain, namely, the excited waves travel freely
along the chain without any energy loss, no temperature gradient can be
set up, and the heat current remains a constant even though the size of
the chain is changed.  To testify this argument, we show the quantity
$J(400)/J(800) $ versus $T=(T_++T_-)/2$ in Fig. 3(d), where $J(N)$ is the
heat current flux for the system of size $N$. In the case of a
size-independent $J(N)$ one should get $J(400)/J(800) = 1$, otherwise one
would get $J(400)/J(800)>1 $.  Fig. 3(d) captures this transition nicely for
the FPU chain.  The corresponding temperature threshold is about
$T_c\sim0.01$. In the region of $T\sim0.001$ the numerical calculations do
show that no temperature gradient is formed. 

The different scattering mechanism in the FPU chain and those chains having
normal thermal conductivity leads to a different temperature dependence of
thermal conductivity $\kappa(T)$. In Fig. 4 we plot $\kappa(T)$ for a FPU
chain with an external potential of form $U(x)=-\gamma\cos(2\pi x)$ for four
different values of $\gamma=0, 0.01, 0.05$ and $0.1$. The chain size is 
fixed at $N=100$. As pointed out above, in small $\gamma$ regime such as 
$\gamma=0$ and $0.01$, the energy transport
is assisted by the solitary waves, the system has a large $\kappa$  
which
decreases as the temperature is increased.  However, in the opposite regime 
($\gamma=0.05$ and $0.1$),
the energy transport is diffusive and obeys the Forier law, $\kappa$
increases with temperature, because more phonons are excited.

\begin{figure}
\epsfxsize=8cm
\epsfbox{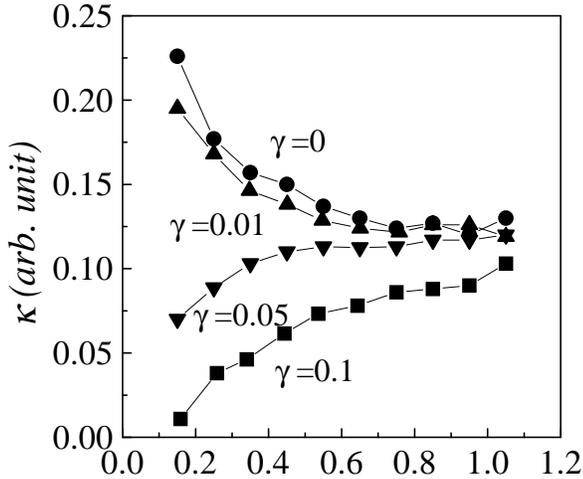}
\vspace{-2.5cm}
\narrowtext
\caption{The thermal conductivity $\kappa (T)$ for the FPU chain with an 
external potential $U(x) = -\gamma\cos(2\pi x)$. 
} 
\end{figure} 

In summary, by studying two classes of 1D nonintegrable lattices, we have
answered the two questions raised at the beginning: (i) The multiple
scattering of the excited modes by the external potential leads to a decay
(at least exponentially) of correlation, so that a diffusive transport
process can be reached, and the heat conduction obeys the Fourier law;
(ii) Although the interaction of solitary waves makes it possible to set
up temperature gradient in the FPU and alike nonintegrable models, the
momentum conservation prohibits the diffusive transport and consequently
leading to the divergent thermal conductivity.  In addition, we have
uncovered an important fact in the FPU model, namely, the existence of a
threshold temperature, below which the FPU mode behaves like a harmonic
chain. 

\bigskip
BL would like to thank G. Casati for useful discussions.
This work was supported in part by Hong 
Kong Research Grant
Council and the Hong Kong Baptist University Faculty
Research Grant.

{\it Note added in proof}. -- After submission of this paper we get to
know the following results. Prosen and Campbell\cite{PC99} proved in a
more rigorous way that for a 1D classical many-body lattice total momentum
conservation implies anomalous conductivity. The normal thermal
conductivity in the $\phi^4$ lattice has also been observed by Aoki and
Kusnezov\cite{AK99}. The role of the external potential has been further
studied by Tsironis {\it et al}\cite{Bishop99}.

\end{multicols}

\begin{thebibliography}{9}

\bibitem{chaos98}
``{\it Focus issue: Chaos and Irreversibility}'', Chaos {\bf 8}, 
(1998); J. Lebowitz, Physica A {\bf 263}, 516 
(1999); I. Prigogine, {\it ibid.}  p. 528; D. Ruelle, {\it 
ibid.}  p. 540.

\bibitem{Lebowitz78}
J. L. Lebowitz and H. Spohn, J. Stat. Phys. {\bf 19}, 633 (1978).

\bibitem{Casati84} 
G. Casati, J. Ford, F. Vivaldi, and W. M. Visscher, 
Phys. Rev. Lett. {\bf 52}, 1861 (1984);
T. Prosen and M. Robnik, J. Phys. A {\bf 25}, 3449 
(1992); D. J. R. Mimnagh and L.E. Ballentine, Phys. Rev. E {\bf 56}, 5332 
(1997); H. A. Posch and Wm. G. Hoover, {\it ibids.} {\bf 58} 4344 (1998).

\bibitem{FPU}
H. Kaburaki and M. Machida, Phys. Lett. A {\bf 181}, 85 
(1993); S. Lepri, R. Livi, and A Politi, Phys. Rev. Lett. {\bf 78},
1897 (1997).

\bibitem{HLZ98} 
B. Hu, B. Li, and H. Zhao, Phys. Rev. E {\bf 57}, 2992 (1998).

\bibitem{FHLZ98} 
A. Fillipov, B. Hu, B. Li, and A. Zeltser, J. Phys. A {\bf 31}, 7719 (1998).

\bibitem{Hatano99}
T. Hatano, Phys. Rev. E {\bf 59}, R1 (1999).

\bibitem{HLZ99}
B. Hu, B. Li, and H. Zhao (to be published)

\bibitem{Dhar99}
A. Dhar and D. Dhar, Phys. Rev. Lett. {\bf 82}, 480 (1999).

\bibitem{Casati99}
D. Alonso, R. Artuso, G. Casati, and I. Guarneri, Phys. Rev. Lett. {\bf 
82}, 1589 (1999).

\bibitem{TLH99}
P.-Q. Tong, B. Li, and B. Hu, Phys. Rev. B {\bf 59}, 8639 (1999).

\bibitem{Rieder}
Z. Rieder, J. L. Lebowitz, and E. Lieb, J. Math. Phys. {\bf 8}, 1073 
(1967).

\bibitem{Aubry96}
D. Chen, S. Aubry, and G. P. Tsironis, Phys. Rev. Lett. {\bf 77}, 4776 
(1996).

\bibitem{NH85} 
S. Nos$\acute{e}$, J. Chem. Phys. {\bf 81}, 511 (1984);  
W. G. Hoover, Phys. Rev. A {\bf 31}, 1695 (1985).

\bibitem{PC99}
T. Prosen and D. K. Campbell, LANL-preprint chao-dyn/9908021.

\bibitem{AK99}
K. Aoki and D. Kusnezov, LANL-Preprint chao-dyn/9910015.

\bibitem{Bishop99}
G. P. Tsironis, A. R. Bishop, A. V. Savin, and A. V. Zolotaryuk, Phys.
Rev. E {\bf 60}, 6610(1999).

\end{thebibliography}
\end{document}